\documentclass[12pt,preprint]{aastex}

\def\etal{{\it et al.} }

\begin{document}
\title{The synchrotron-self-Compton radiation
accompanying shallow decaying X-ray afterglow: the case of GRB
940217}

\author{Da-Ming Wei$^{1,2,3}$ and Yi-Zhong Fan$^{1,2,4,5}$}
\affil{$^1$ Purple Mountain Observatory, Chinese Academy
of Science, Nanjing 210008, China.\\
$^2$ National Astronomical Observatories,Chinese Academy of
Sciences, Beijing, 100012, China.\\
$^3$ Joint Center for Particle Nuclear Physics and Cosmology of
Purple Mountain Observatory - Nanjing University, Nanjing
210008,China.\\
$^4$ The Racah Inst. of Physics, Hebrew University, Jerusalem
91904, Israel.\\
$^5$ Golda Meir Fellow.\\
}

\begin{abstract}
High energy emission ($>$ tens MeV) of Gamma-Ray Bursts (GRBs)
provides an important clue to understand the physical processes
involved in GRBs, which may be correlated with the GRB early
afterglow. A shallow decline phase has been well detected in about
half {\it Swift} Gamma-ray Burst X-ray afterglows. The widely
considered interpretation involves a significant energy injection
and possibly time-evolving shock parameter(s). This work we
calculate the synchrotron-self-Compton (SSC) radiation of such an
external forward shock and show that it could explain the well-known
long term high energy (i.e., tens MeV to GeV) afterglow of GRB
940217. We propose that the cooperation of Swift and GLAST will help
to reveal the nature of GRBs.
\end{abstract}


\section{Introduction}
\label{sect:Intro} Among the high energy (above tens MeV)
afterglow of Gamma-ray Burst (GRB) detected so far, that of GRB
940217 is the longest and also the most energetic one. The sub-GeV
emission lasted more than 5000 seconds and it included also a 18
GeV photon \cite{Hurley94}. The spectrum in the energy range
$1~{\rm MeV}$ to 18 GeV, cannot be fitted with a simple power law
(see Fig. 3 of Hurley et al. 1994). A new spectral component in
the energy range larger than several tens MeV is needed. This
finding motivates many interesting ideas: (i) the interaction of
ultra-relativistic protons with a dense cloud \cite{Katz94}, (ii)
synchrotron-self-Compton (SSC) scattering in early forward/reverse
shocks or during the prompt emission \cite{MR94} and (iii) an
electromagnetic cascade of TeV $\gamma-$rays in the
infrared/microwave background \cite{Plaga95}. However, the two
important observation facts: (a) the count rate of high energy
photons is almost a constant (b) the typical energy of these
photons is nearly unchanged, have not been satisfactorily
reproduced.

To interpret the unusual high energy afterglow of GRB 940217, the
physical processes involved in the early GRB afterglow phase is
highly needed. The successful launch of the {\it Swift} satellite
did open a new window to reveal what happens in the early GRB
afterglow phase. As summarized in Zhang et al. (2006) and Nousek
et al. (2006), in a canonical {\it Swift} GRB X-ray afterglow
lightcurve some interesting features are emerging, including the
very early sharp decline (i.e., phase-I), a shallow decline of the
X-ray afterglow (i.e., phase-II), and the energetic X-ray flares
(i.e., phase-V). The interpretation and the implication of these
features have been discussed in great detail (see M\'esz\'aros
2006; Piran \& Fan 2007; Zhang 2007 for recent reviews). For
Phase-II, which interests us here, a widely considered explanation
is a significant energy injection $dE_{\rm inj}/dt \propto t^{-q}$
(see Zhang et al. 2006 and Nousek et al. 2006 and the references
therein). However, the energy injection process, if there is,
seems to be not all the story. As shown in Fan \& Piran (2006),
for some GRBs with good quality multi-wavelength afterglow data,
the X-ray and optical light curves break chromatically and thus
challenge the energy injection model (see also Panaitescu \etal
2006, Huang \etal. 2007). An assumption additional to the energy
injection to solve such a puzzle is that the shock parameter(s)
may be shock strength dependent (i.e., time-dependent).

In this work, we calculate the synchrotron-self-Compton (SSC)
radiation of the external forward shock undergoing a significant
energy injection. The shock parameters, $\epsilon_e$ and
$\epsilon_B$, the fraction of shock energy given to the shocked
electrons and magnetic field, are assumed to be time-dependent. We
show the high energy afterglow of GRB 940217 could be given rise to
in such a scenario.

\section{The SSC emission of the forward shock}

In this section we calculate the synchrotron-self-Compton (SSC)
radiation of the external forward shock undergoing a significant
energy injection and evolving shock parameters.

\subsection{The SSC emission in the standard fireball model}

{\bf Synchrotron radiation.} As usual, we assume that the
electrons accelerated by the shock would follow the power law
distribution $dn_e/d\gamma_e \propto \gamma_e^{-p}$ for $\gamma_e
> \gamma_{m}$, where $\gamma_{m}$ is the minimum Lorentz
factor of shocked electrons (Sari et al. 1998). Then following Yost
\etal (2003) and Fan \& Piran (2006), the observed typical frequency
of synchrotron radiation is $\nu_m=4.2\times
10^{14}(\frac{1+z}{2})^{1/2}\epsilon_{e,-1}^2
\epsilon_{B,-2}^{1/2}E_{52}^{1/2}t_3^{-3/2} ~{\rm Hz}$. Where $z$ is
the redshift of the GRB, $\epsilon_e$ and $\epsilon_B$ are the
energy fraction occupied by the electrons and magnetic field
respectively, $E$ is the isotropic energy, $t$ is the observer time.
Here the convention $Q_x=Q/10^x$ has been adopted in cgs units
throughout the text, and we have taken the spectral index of the
electron distribution $p=2.5$. The observed cooling frequency is
$\nu_c=4.1\times 10^{16}(\frac{1+z}{2})^{-1\over 2}
\epsilon_{B,-2}^{-3\over 2}E_{52}^{-1/2}n^{-1}t_3^{-1/2}(1+Y)^{-2}
~{\rm Hz}$, where $n$ is the surrounding medium density,
$Y=[-1+\sqrt{1+4x\epsilon_e/\epsilon_B}]/2$ is the Compton
parameter, $x\simeq min\{1, (\nu_m/\nu_c)^{(p-2)/2}\}$ (Sari \& Esin
2001). The peak flux is $ F_{\nu,max}=3.2(\frac{1+z}{2})
\epsilon_{B,-2}^{1/2}E_{52}n^{1/2}D_{28}^{-2} ~{\rm mJy}$. Where
$D_{28}$ is the luminosity distance in units of $10^{28}$ cm. So in
the fast cooling phase ($\nu_c < \nu_m$) the light curve is (Sari
\etal 1998): $F_\nu \propto t^{1/6}(1+Y)^{2/3}$ for $\nu_c>\nu$,
$F_\nu \propto t^{-1/4}(1+Y)^{-1}$ for $\nu_m>\nu>\nu_c$, and $F_\nu
\propto t^{-(3p-2)/4}(1+Y)^{-1}$ for $\nu>\nu_m$. While in the slow
cooling phase ($\nu_c > \nu_m$) the light curve is:  $F_\nu \propto
t^{1/2}$ for $\nu_m>\nu$, $F_\nu \propto t^{-3(p-1)/4}$ for
$\nu_c>\nu>\nu_m$, and $F_\nu \propto t^{-(3p-2)/4}(1+Y)^{-1}$ for
$\nu>\nu_c$.

Please note that in previous works the evolution of the Compton
parameter $Y$ has always been ignored. However, If $Y\gg 1$, then
the effect of $Y$ evolution should be considered. From the relation
$x= min\{1, (\nu_m/\nu_c)^{(p-2)/2}\}$ we know that, in the fast
cooling phase, $x=1$, then $Y$ is independent of the time, while in
the slow cooling phase $Y$ should evolve with time.
$x=(\nu_c/\nu_m)^{-(p-2)/2}\propto (1+Y)^{p-2}t^{-(p-2)/2}$, then we
get

\begin{equation}
Y\simeq
10^{\frac{3-p}{4-p}}(\frac{1+z}{2})^{\frac{p-2}{2(4-p)}}\epsilon_{e,-1}^{\frac{p-1}{4-p}}
\epsilon_{B,-2}^{\frac{p-3}{4-p}}E_{52}^{\frac{p-2}{2(4-p)}}n^{\frac{p-2}{2(4-p)}}
t_3^{-\frac{p-2}{2(4-p)}}
\end{equation}

So we find $Y\propto t^{-(p-2)/2(4-p)}$, then $\nu_c\propto
t^{(3p-8)/2(4-p)}$, and so we have $F_{\nu}\propto
t^{\frac{3p^2-12p+4}{4(4-p)}}$ for $\nu>\nu_c$.

{\bf The SSC emission.} The effect of SSC process on GRB afterglow
emission has been discussed by several authors (e.g. Wei \& Lu
1998, 2000; Sari \& Esin 2001). The typical frequency of SSC
emission is
\begin{eqnarray}
\nu_m^{IC}&=&2\gamma_m^2\nu_m=3.36\times
10^{21}(\frac{1+z}{2})^{1/2}\nonumber\\
&&\epsilon_{e,-1}^4
\epsilon_{B,-2}^{1/2}E_{52}^{3/4}n^{-1/4}t_3^{-9/4} ~{\rm Hz}
\end{eqnarray}
Where $\gamma_m=\epsilon_e(\frac{p-2}{p-1})\frac{m_p}{m_e}\Gamma
\simeq 2\times
10^{3}\epsilon_{e,-1}E_{52}^{1/8}n^{-1/8}t_3^{-3/8}$ is the
minimum Lorentz factor of shocked electrons, $\Gamma$ is the bulk
Lorentz factor, the cooling Lorentz factor $\gamma_c \simeq
2\times
10^{4}\epsilon_{B,-2}^{-1}E_{52}^{-3/8}n^{-5/8}(\frac{1+z}{2})^{-1/2}t_3^{1/8}(1+Y)^{-1}$,
then the cooling frequency of SSC emission is
\begin{eqnarray}
\nu_c^{IC} &\approx & 2\gamma_c^2\nu_c =3.28\times
10^{25}(\frac{1+z}{2})^{-3/2} \nonumber\\ &&
\epsilon_{B,-2}^{-7/2}E_{52}^{-5/4}n^{-9/4}t_3^{-1/4}(1+Y)^{-4}
~{\rm Hz}
\end{eqnarray}

The peak flux of the SSC emission is (Sari \& Esin 2001)
\begin{eqnarray}
F_m^{IC} &\simeq & \frac{1}{3}n\sigma_T rF_{\nu,max}=1.2\times
10^{-13}(\frac{1+z}{2})\nonumber\\ &&
\epsilon_{B,-2}^{1/2}E_{52}^{5/4}n^{5/4}t_3^{1/4}D_{28}^{-2} ~{\rm
erg cm^{-2}s^{-1}MeV^{-1}}
\end{eqnarray}

Therefore, in the fast cooling phase, the light curve is
\begin{equation}
F_\nu \propto \cases{
t^{1/3}(1+Y)^{4/3}, & $\nu_c^{IC}>\nu$, \cr
t^{1/8}(1+Y)^{-2}, & $\nu_m^{IC}>\nu>\nu_c^{IC}$, \cr
t^{-(9p-10)/8}(1+Y)^{-2}, & $\nu>\nu_m^{IC}$, \cr }
\end{equation}

and in the slow cooling phase, the light curve is
\begin{equation}
F_\nu \propto \cases{
t, & $\nu_m^{IC}>\nu$, \cr
t^{-(9p-11)/8}, & $\nu_c^{IC}>\nu>\nu_m^{IC}$, \cr
t^{-(9p-10)/8}(1+Y)^{-2}, & $\nu>\nu_c^{IC}$. \cr }
\end{equation}

Again, if considering $Y$ evolution, then, $F_\nu\propto
t^{(9p^2-38p+24)/8(4-p)}$ for $\nu>\nu_c^{IC}$.

\subsection{The SSC emission with energy injection and evolving shock parameters}

{\bf A. The SSC emission with energy injection.} In the standard
fireball model, the shock energy $E$ is assumed to be constant.
However, there are increasing evidences that the shock energy may
increase with time during some period. One good example is the
discovery of the "shallow decay phase" in the early X-ray light
curves of many GRBs, which is usually attributed to the energy
injection (Zhang \etal 2006; Nousek \etal 2006). Now we consider
the case that there is significant continuous energy injection
into the fireball, so the fireball decelerates less rapidly and
the afterglow emission will show a shallow decline. The dynamical
evolution and the synchrotron radiation signature for energy
injection have been discussed by many authors (Rees \&
M\'esz\'aros 1998; Dai \& Lu 1998; Cohen \& Piran 1999; Zhang \&
M\'esz\'aros 2001a; Zhang \etal 2006; Nousek \etal 2006; Fan \& Xu
2006).

Here we assume the energy injection takes a form $dE_{\rm inj}/dt
\propto t^{-q}$ \cite{Cohen99,ZM01}, then the fireball energy
evolves with time as $E\propto t^{1-q}$, the fireball radius
$r\propto t^{(2-q)/4}$, the minimum electron Lorentz factor
$\gamma_m\propto \Gamma\propto t^{-(2+q)/8}$, the observed typical
frequency of synchrotron radiation $\nu_m\propto \gamma_m^2
B\Gamma\propto t^{-(2+q)/2}$, the cooling Lorentz factor
$\gamma_c\propto (1+Y)^{-1}t^{(3q-2)/8}$, and the observed cooling
frequency of synchrotron radiation $\nu_c\propto
\gamma_c^{2}B\Gamma\propto (1+Y)^{-2}t^{-(2-q)/2}$, the peak flux
$F_{\nu, max}\propto N_e B\Gamma\propto t^{1-q}$.  The typical
frequency of SSC emission is $\nu_m^{IC}\propto\gamma_m^2
\nu_m\propto t^{-3(2+q)/4}$, the cooling frequency of SSC emission
is $\nu_c^{IC}\propto \gamma_c^2 \nu_c\propto
(1+Y)^{-4}t^{(5q-6)/4}$, and the peak flux of SSC radiation is
$F_m^{IC}\simeq \frac{1}{3}n\sigma_T rF_{\nu,max}\propto
t^{(6-5q)/4}$. Using these relations, we can get the light curves
of synchrotron radiation and SSC emission with energy injection.

{\bf B. The SSC emission with evolving shock parameters.} In the
standard fireball model, the shock parameters $\epsilon_e$ and
$\epsilon_B$ are assumed to be constant. However, it is also
possible that these quantities may vary with time. Yost \etal
(2003), Fan \& Piran (2006) and Ioka \etal (2006) have discussed the
afterglow emission with $\epsilon_e$ and $\epsilon_B$ being
time-dependent. By modeling the afterglow of several GRBs, it was
found that the values of $\epsilon_e$ and $\epsilon_B$ of the
forward shock are quite different from that of reverse shock (Fan
\etal 2002; Zhang, Kobayashi \& M\'esz\'aros 2003; Kumar \&
Panaitescu 2003; Wei \etal 2006). We note that the forward shock is
ultra-relativistic, while the reverse shock is mild-relativistic, so
this result suggests that the shock parameters may be correlated
with the strength of the shock.

Following Fan \& Piran (2006), here we simply assume $\epsilon_e
\propto \Gamma^{-a}$, $\epsilon_B \propto \Gamma^{-b}$, and since
$\Gamma \propto t^{-(2+q)/8}$, so $\epsilon_e \propto t^{(2+q)a/8}$,
$\epsilon_B \propto t^{(2+q)b/8}$. Then we can obtain the light
curve of synchrotron radiation and SSC emission.

{\bf C. The SSC emission with both the above effects.} It is also
possible that during the shock evolution, both the shock energy and
the shock parameters evolve with time. Based on the previous
analysis, we can obtain the synchrotron radiation and SSC emission
light curves easily. Table 1 gives the temporal index $\alpha$ of
the afterglow emission, where $F_\nu \propto t^{-\alpha}$ is
adopted. We define $\alpha=\alpha_0+\alpha_E+\alpha_v+\alpha_Y$,
where $\alpha_0$ corresponds to the contribution of the standard
emission, $\alpha_E$ represents the contribution of the energy
injection, $\alpha_v$ stands for the contribution of evolving shock
parameters, and $\alpha_Y$ comes from the evolution of Compton
parameter Y. For example, if we only consider energy injection, then
$\alpha=\alpha_0+\alpha_E$. If only evolving shock parameters is
considered, then $\alpha=\alpha_0+\alpha_v$. If both the effects are
considered, then $\alpha=\alpha_0+\alpha_E+\alpha_v$. If $Y\gg 1$,
then the term $\alpha_Y$ should be included.

\section{The case of GRB940217}

GRB940217 was a very famous burst for its long-lasting high energy
afterglow emission (Hurley \etal 1994). The high energy photons
($E>30$ MeV) were recorded for about 5400 seconds, including an 18
GeV photon $\sim$ 4500s after the low energy emission had ended.
The 30 MeV to 30 GeV EGRET spectrum is $(1.3\pm 0.4)\times
10^{-8}(E/86MeV)^{-2.83\pm 0.64} {\rm photons}~ {\rm cm}^{-2}{\rm
s}^{-1}{\rm MeV}^{-1}$, excluding the 18 GeV photon. By
integrating this spectrum, the fluence at $>30$ MeV is $7\times
10^{-6} {\rm erg~cm^{-2}}$. In addition, for this GRB there are
two important observation facts: (a) the count rate of high energy
photons is almost a constant; (b) the typical energy of these
photons is nearly unchanged. This two facts imply that the flux
should be nearly a constant $\sim 1.4\times 10^{-9}{\rm
erg~cm^{-2}~s^{-1}}$.

Based on the above analysis, if we assume that the energy injection
occurred at time $t \sim$ 500s, and we take the parameters as
follows: $\epsilon_{\rm e,-1}\sim 0.7$, $\epsilon_{\rm B,-2}\sim
0.5$, $n\sim 1$, $E_{52}\sim 5$, $z\sim 0.1$, then at this time, the
typical frequency of SSC emission is $\nu_m^{IC} \sim 20$ MeV, the
cooling frequency of SSC emission is $\nu_c^{IC} \sim 60$ GeV, and
the peak flux of SSC emission is $F_m^{IC} \sim 1.4\times
10^{-11}~{\rm ergs}~ {\rm cm}^{-2}{\rm s}^{-1}{\rm MeV}^{-1}$, which
is quite agreement with the observation. In addition, we note that
the observed photon energies lie between $\nu_m^{IC}$ and
$\nu_c^{IC}$, and since the spectrum is very soft, $\beta=1.83\pm
0.64$, so the index of electron distribution is $p \sim 4$. From
table 1 we can find that, in the standard case, the flux would
decrease with time as $t^{-25/8}$, which is obviously inconsistent
with the observation. If we consider energy injection, then
$F_{\nu}\propto t^{-(6+19q)/8}$, since $0<q<1$, so the flux would
decay steeper than $t^{-3/4}$, which is still inconsistent with the
observation. Then, there is only one choice - the shock parameters
should evolve with time. If we take $q\sim 0.5$ (which is suggested
by the recent {\em Swift} XRT data, see Zhang \etal 2006), $a\sim
1$, $b \sim 0$, then the flux would be nearly a constant, which is
agreement with the observation.

In order to investigate the SSC emission more carefully, Fan \etal
(2007) have developed a numerical code to calculate the Compton
process self-consistently for GRB high energy afterglow emission.
Using this code, we calculate the SSC emission numerically, the
result has been shown in Fig.1, the parameter are: the initial
kinetic energy is $4\times 10^{52}$ erg, $n=1$, $z=0.1$,
$\epsilon_{\rm B,-2}=0.3$, $p=4$, and $\theta_j=0.2$. For $10~{\rm
s}<t<5000~{\rm s}$, the energy injection form is taken to be
$dE/{dt}=4\times 10^{50}(t/10{\rm s})^{-0.55}$ and
$\epsilon_e=0.06(t/5000 {\rm s})^{0.4}$. At late times, the energy
injection disappears and $\epsilon_e=0.06$.  From Fig.1 we find
that the numerical results are consistent with the analytic
estimates and can well account for the observation (both the light
curve and the spectrum) of GRB940217.

\section{Discussion and conclusion}
Since its discovery, the long-lasting high energy afterglow emission
of GRB 940217 has been extensively discussed (e.g. Katz 1994;
M\'esz\'aros \& Rees 1994; Plaga 1995; Cheng \& Cheng 1996; Dermer
\etal 2000; Wang et al. 2001; Zhang \& M\'esz\'aros 2001b; Guetta \&
Granot 2003; Pe'er \& Waxman 2004). However, we note that the SSC
emission of the standard afterglow model cannot well account for the
observation of GRB940217. For example, Dermer et al. (2000)
calculated the synchrotron-self-Compton emission during the blast
wave propagation, but their spectrum is very hard, which is
inconsistent with the observation of GRB 940217. As shown in Tab.1,
the flux of SSC emission would decrease with time as $t^{-25/8}$ for
a $p\sim 4$ that is needed to reproduce the very steep MeV to GeV
spectrum, even when considering the energy injection, the flux would
still decay steeper than $t^{-3/4}$. Therefore, the nearly constant
of the high energy flux strongly suggests that other physical
processes should be involved, such as the evolution of shock
parameters and/or energy injection considered in this paper. Some
possible energy injection processes have been proposed in Rees \&
M\'esz\'aros (1998) and Dai \& Lu (1998), while the possible
physical scenario giving rise to the time-evolving shock parameters
are far from clear \cite{PF06}. The peculiar chromatic break
detected in quite a few early X-ray and optical afterglow data
\cite{FP06a,Panaitescu06}, however, did indicate such a possibility.

In previous analysis, the dependence of the Compton parameter $Y$
on time has always been ignored. However, from table 1 we can see
that, in some cases, the influence of $Y$ evolution cannot be
ignored. For example, in the standard case ($q=1$, $a=b=0$, i.e.
without energy injection and the shock parameters are constant),
the light curves of synchrotron radiation ($\nu>\nu_c$) and SSC
emission ($\nu>\nu_c^{IC}$) will be flattened by $t^{1/6}$ and
$t^{1/3}$ respectively for $p=2.5$. If $p=3$, then the effect will
be more prominent, the light curves of synchrotron radiation
($\nu>\nu_c$) and SSC emission ($\nu>\nu_c^{IC}$) will be
flattened by $t^{1/2}$ and $t^{1}$ respectively. So under some
circumstances, the effect of $Y$ evolution should be considered.

GLAST will be lunched soon, and it is expected that GLAST will
detect high energy emission (20 MeV to 300 GeV) of GRBs, which may
open a new window to understand the physical processes occurred in
GRBs. We hope that GLAST can detect more events like GRB940217,
and this can provide important clues to explore the nature of GRBs
(see Fan \etal 2007 for extensive discussions).

\section*{Acknowledgments}

This work is supported by the National Natural Science Foundation
(grants 10225314, 10233010, 10621303 and 10673034) of China.

\clearpage

\begin{table}
\caption{The Temporal index $\alpha$ of afterglow emission. Here
$F_\nu \propto t^{-\alpha}$ is adopted. We define
$\alpha=\alpha_0+\alpha_E+\alpha_v+\alpha_Y$, where $\alpha_0$
corresponds to the contribution of standard emission, $\alpha_E$
represents the contribution of energy injection, $\alpha_v$ stands
for the contribution of evolving shock parameters, and $\alpha_Y$
comes from the evolution of Compton parameter Y.\label{Table1}}
\begin{tabular}{lllll}
\hline\hline

& $\alpha_0$ & $\alpha_E $  &  $\alpha_v $ & $\alpha_Y$ \\
\hline
Synchrotron radiation & slow cooling \\
\hline
$\nu<\nu_m$   &  $-{1 \over 2}$  &   $-{5(1-q)\over 6}$ & $ {(2a-b)(2+q) \over 24}$ & $0$ \\
$\nu_m<\nu<\nu_c$   &  ${3(p-1)\over 4}$  &  $-{(1-q)(p+3)\over 4}$
  &  $-{4a(2+q)(p-1)+b(2+q)(p+1) \over 32}$ & $0$ \\
$\nu>\nu_c$   &  ${3p-2 \over 4}$  &   $-{(1-q)(p+2) \over 4}$  & $-{4a(2+q)(p-1)+b(2+q)(p-2) \over 32}$ & $-{4q(p-2)-a(2+q)(p-1)-b(2+q)(p-3) \over 8(4-p)}$ \\

\hline
Synchrotron radiation & fast cooling \\
\hline
$\nu<\nu_c$   &  $-{1\over 6}$  &   $-{7(1-q) \over 6}$ &  $-{(2+q)b \over 8}$  &  $-{(2+q)(a-b) \over 24}$ \\
$\nu_c<\nu<\nu_m$   &  ${1\over 4}$  &  $-{3(1-q)\over 4}$  & ${(2+q)b \over 32}$ &  ${(2+q)(a-b) \over 16}$ \\
$\nu>\nu_m$   &  ${3p-2 \over 4}$ &   $-{(1-q)(p+2) \over 4}$  &  $-{4a(2+q)(p-1)+b(2+q)(p-2) \over 32}$ &  ${(2+q)(a-b) \over 16}$ \\

\hline
SSC emission & slow cooling \\
\hline
$\nu<\nu_m^{IC}$   &  $-1$   & $-(1-q)$  &  ${(2+q)(4a-b) \over 24}$ & $0$ \\
$\nu_m^{IC}<\nu<\nu_c^{IC}$   &  ${9p-11 \over 8}$  &   $-{(1-q)(3p+7) \over 8}$  &   $-{8a(2+q)(p-1)+b(2+q)(p+1) \over 32}$ & $0$ \\
$\nu>\nu_c^{IC}$   &  ${9p-10 \over 8}$  &   $-{(1-q)(3p+2) \over 8}$  &  $-{8a(2+q)(p-1)-b(2+q)(6-p) \over 32}$   &   $-{4q(p-2)-a(2+q)(p-1)-b(2+q)(p-3) \over 4(4-p)}$ \\

\hline
SSC emission & fast cooling  \\
\hline
$\nu<\nu_c^{IC}$   &  $-{1 \over 3}$   &   $-{5(1-q) \over 3}$ & $-{5b(2+q) \over 24}$ &  $-{(2+q)(a-b) \over 12}$ \\
$\nu_c^{IC}<\nu<\nu_m^{IC}$   &  $-{1 \over 8}$  &   $-{5(1-q) \over 8}$  & ${5b(2+q) \over 32}$  & ${(2+q)(a-b) \over 8}$ \\
$\nu>\nu_m^{IC}$   &  ${9p-10 \over 8}$  &  $-{(1-q)(3p+2) \over 8}$ & $-{8a(2+q)(p-1)-b(2+q)(6-p) \over 32}$ &  ${(2+q)(a-b) \over 8}$ \\
\hline
\end{tabular}
\label{Tab:alpha-beta}
\end{table}

\clearpage

\begin{figure}
\plotone{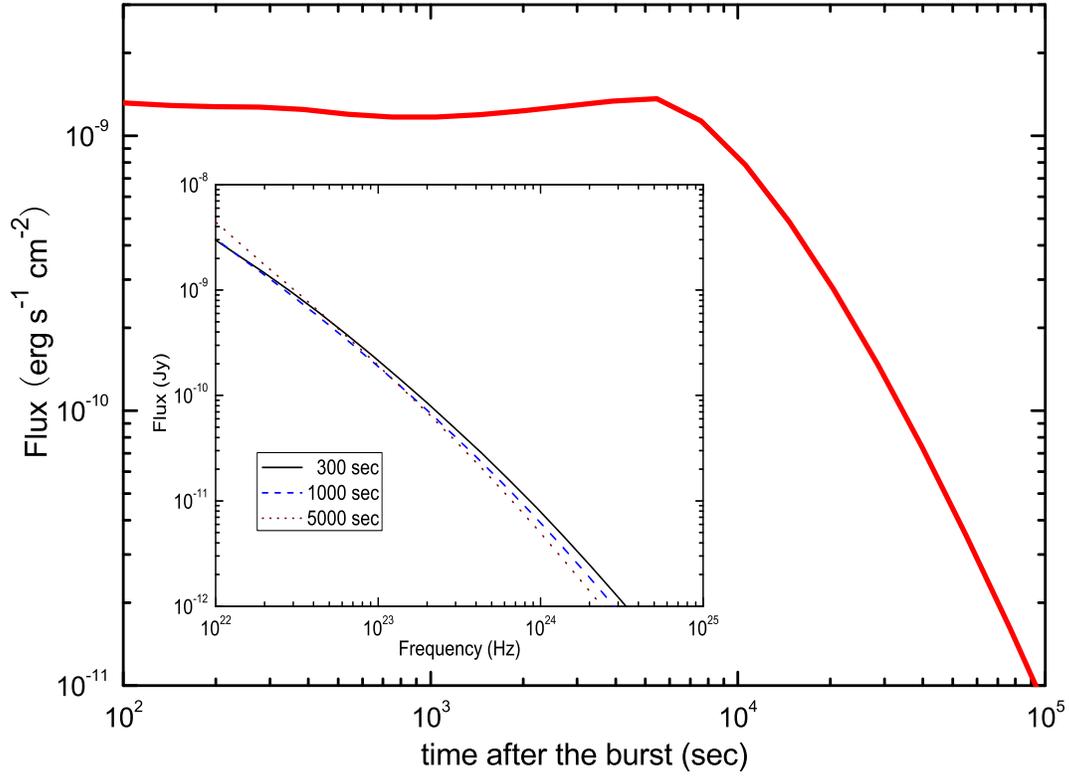}
\caption{The SSC radiation of the forward shock
undergoing energy injection and with evolving shock parameters, the
case of GRB 940217: the thick solid line is the light curve and the
inserted plot is the spectrum (the times have been marked in the
plot), in the energy range of 30 MeV - 30 GeV. For $10~{\rm
s}<t<5000~{\rm s}$, the energy injection form is taken to be
$dE/{dt}=4\times 10^{50}(t/10{\rm s})^{-0.55}$ and
$\epsilon_e=0.06(t/5000 {\rm s})^{0.4}$. At late times, the energy
injection disappears and $\epsilon_e=0.06$. Other parameters
involved in the calculation are as follows: the initial kinetic
energy is $4\times 10^{52}$ erg, $z=0.1$, $\epsilon_{\rm B,-2}=0.3$,
$p=4$, and $\theta_j=0.2$.} \label{Fig:fig1}
\end{figure}


\begin{thebibliography}{99}
\bibitem[Cheng \& Cheng 1996]{CC96} Cheng, L. X. \& Cheng, K. S.,
1996, ApJ, 459, L79
\bibitem[Cohen \& Piran 1999]{Cohen99}
Cohen, E., \& Piran, T., 1999, ApJ, 518, 346
\bibitem[]{517} Dai, Z.G., \& Lu, T., 1998, A\&A, 333, L87
\bibitem[Dermer \etal 2000]{Dermer00} Dermer, C.D., Chiang, J., \&
Mitman, K., 2000, \apj, 537, 785
\bibitem[Fan \etal 2007]{Fan07} Fan,Y.Z. \etal, 2007, \mnras, to be
submitted
\bibitem[Fan \etal 2002]{Fan02} Fan, Y.Z., Dai, Z.G., Huang, Y.F., \& Lu, T., 2002,
Chin. J. Astron. Astrophys., 2, 449
\bibitem[Fan \& Piran 2006]{FP06a} Fan,Y.Z., \& Piran, T. 2006, \mnras, 369, 197
\bibitem[Fan \& Xu 2006]{FX06} Fan,Y.Z., \& Xu, D. 2006, \mnras, 372, L19
\bibitem[Guetta \& Granot 2003]{GG03} Guetta, D.,  \& Granot, J., 2003, \apj, 585, 885
\bibitem[Huang \etal 2007]{Huang06} Huang, K. Y., et al., ApJ,
654, L25
\bibitem[Hurley \etal 1994]{Hurley94} Hurley K., \etal, 1994, \nat, 372, 652
\bibitem[]{528} Ioka, K., Toma, K., Yamazaki, R., Nakamura, T., 2006,
A\&A, 458, 7
\bibitem[Katz 1994]{Katz94} Katz, J. I., 1994, \apj, 432, L27
\bibitem[]{531} Kumar, P., \& Panaitescu, A., 2003, MNRAS, 346, 905
\bibitem[M\'esz\'aros 2006]{Meszaros06}{M\'esz\'aros, P.}, 2006, {Rep. Prog. Phys}, {69}, {2259}
\bibitem[M\'{e}sz\'{a}ros \& Rees 1994]{MR94} M\'{e}sz\'{a}ros, P., \& Rees, M. J., 1994, \mnras, 269,
L41
\bibitem[Nousek \etal 2006]{Nousk06} {Nousek, J. A. et al.} {2006}, {\apj}, {642}, {389}
\bibitem[Panaitescu \etal 2006]{Panaitescu06} Panaitescu, A. \etal, 2006, \mnras, 369,
2059
\bibitem[]{} Pe'er, A., \& Waxman, E., 2004, \apj, 603, L1
\bibitem[Plaga 1995]{Plaga95} Plaga, R., 1995, \nat, 374, 430
\bibitem[Piran \& Fan 2007]{PF06} Piran, T., \& Fan, Y. Z., 2007, Phil. Trans. R. Soc.
A., in press
\bibitem[Rees \& M\'esz\'aros 1998]{Rees98} Rees, M. J., \& M\'esz\'aros, P., 1998, ApJ, 496, L1
\bibitem[]{542} Sari, R., \& M\'esz\'aros, P., 2000, \apj,535, L33
\bibitem[Sari \& Esin 2001]{Sari01} Sari,R., \& Esin,A.A. 2001, \apj, 548, 787
\bibitem[Sari \etal 1998]{Sari98} Sari,R., Piran,T., \& Narayan,R. 1998, \apj, 497, L17
\bibitem[]{545} Schaefer, B.E., \etal, 1998, \apj, 492, 696
\bibitem[]{546} Schneid, E.J., \etal, 1992, A\&A, 255, L13
\bibitem[]{547} Sommer, M., \etal, 1994, \apj, 422, L63
\bibitem[]{548} Wang, X.Y., Dai, Z.G., \& Lu, T., 2001, \apj, 556, 1010
\bibitem[Wei \& Lu 1998]{Wei98} Wei, D. M., \& Lu,T., 1998, \apj, 505, 252
\bibitem[]{552} Wei, D. M., \& Lu,T., 2000, A\&A, 360, L13
\bibitem[]{553} Wei, D.M., Yan, T., Fan, Y.Z., 2006, \apj, 636, L29
\bibitem[Yost \etal 2003]{Yost03} Yost,S.A., Harrison,F.A., Sari,R., \& Frail,D.A. 2003, \apj, 597, 459
\bibitem[Zhang \& M\'esz\'aros 2001]{ZM01} Zhang, B.,  M\'esz\'aros, P., 2001a, ApJ, 552, L35
\bibitem[]{557} Zhang, B.,  M\'esz\'aros, P., 2001b, ApJ, 559, 110
\bibitem[]{558} Zhang, B., Kobayashi, S., \& M\'esz\'aros, P., 2003,
\apj, 595, 950
\bibitem[Zhang \etal 2006]{Zhang06} Zhang,B., Fan, Y. Z., Dyks, J., Kobayashi, S., M\'esz\'aros, P., \etal 2006, \apj, 642, 354
\bibitem[]{} Zhang, B., 2007, Chin. J. Astron. Astrophys., 7, 1
\end{thebibliography}
\end{document}